\documentclass[useAMS,usenatbib]{mn2e}
\usepackage{graphics}
\usepackage{psfig}
\usepackage{amssymb}
\usepackage{natbib}
\usepackage{times}
\usepackage[dvips]{graphicx}

\begin {document}

\title[Mergers and Halo Spins] 
      {Do Mergers Spin Up Dark Matter Halos?}
\author[Elena D'Onghia \& Julio F. Navarro]
       {Elena~D'Onghia$^{1}$\thanks{Marie Curie Fellow; Email: elena@physik.unizh.ch} \& Julio F. Navarro$^{2}$\thanks{CIAR and Guggenheim Fellow} \\	   
$^1$ Institute for Theoretical Physics, University of Zurich, Winterthurerstrasse 190, CH-8057 Zurich, Switzerland \\
$^2$ Department of Physics and Astronomy, University of Victoria, 3800 Finnerty Road, Victoria, BC, Canada V8P 5C2\\}	       
	     
\date{submitted to MNRAS}
	     
\pagerange{\pageref{firstpage}--\pageref{lastpage}}\pubyear{200?}
\maketitle	     

\label{firstpage}
	     
\begin{abstract}
We use a large cosmological N-body simulation to study the origin of
possible correlations between the merging history and spin of cold
dark matter halos. In particular, we examine claims that remnants of
major mergers tend to have higher-than-average spins, and find that
the effect is driven largely by unrelaxed systems: equilibrium dark
matter halos show no significant correlation between spin and merging
history. Out-of-equilibrium halos have, on average, higher spin than
relaxed systems, suggesting that the virialization process leads to a
net decrease in the value of the spin parameter. We find that this
decrease is due to the internal redistribution of mass and angular
momentum that occurs during virialization. This process is especially
efficient during major mergers, when high angular momentum material is
pushed beyond the virial radius of the remnant. Since such
redistribution likely affects the angular momentum of baryons and dark
matter unevenly, our findings question the common practice of
identifying the specific angular momentum content of a halo with that
of its embedded luminous component. Further work is needed to
elucidate the true relation between the angular momentum content of
baryons and dark matter in galaxy systems assembled hierarchically.
\end{abstract}


\setcounter{footnote}{1}

\section{Introduction}
\label{sec:intro}

Angular momentum has long been regarded as a crucial ingredient of the
cosmic recipe that governs the formation and evolution of galaxies. In
the current structure formation paradigm, the net spin of a dark
matter halo and of its luminous component originates in torques
exerted by neighbouring structures at early times. This tidal-torque
theory, proposed by Hoyle (1949) and developed by Peebles (1969),
Doroshkevich (1970), and White (1984), envisions the spin of a halo
as acquired during the early expansion phase, when tides are strong
and when the moment of inertia that couples the material destined to
collapse to external tides is large.

The acquisition of angular momentum abates after turnaround, as the
moment of inertia of the collapsing material decreases and the
universal expansion pushes the neighbouring matter responsible for
tides ever farther, effectively arresting any further increase in net
angular momentum. In the absence of dissipation, and for reasonably
isolated systems, energy and angular momentum are conserved during the
subsequent collapse and virialization, so the spin of a dark matter
halo is effectively set at turnaround and should evolve little
thereafter (see Porciani et al. 2002a,b for a review of this and more
recent work).

This scenario has interesting and testable consequences. Tidal torques
generate net angular momentum principally through the misalignment
between the gravitational (tidal) shear tensor and the inertia tensor
of the material being spun up. These misalignments are generally weak,
making the spin-up process rather inefficient and leading to a fairly
broad distribution of halo spins that peaks at values well below those
needed for substantial centrifugal support (Porciani et al. 2002b).

\begin{figure}
\includegraphics[width=85mm]{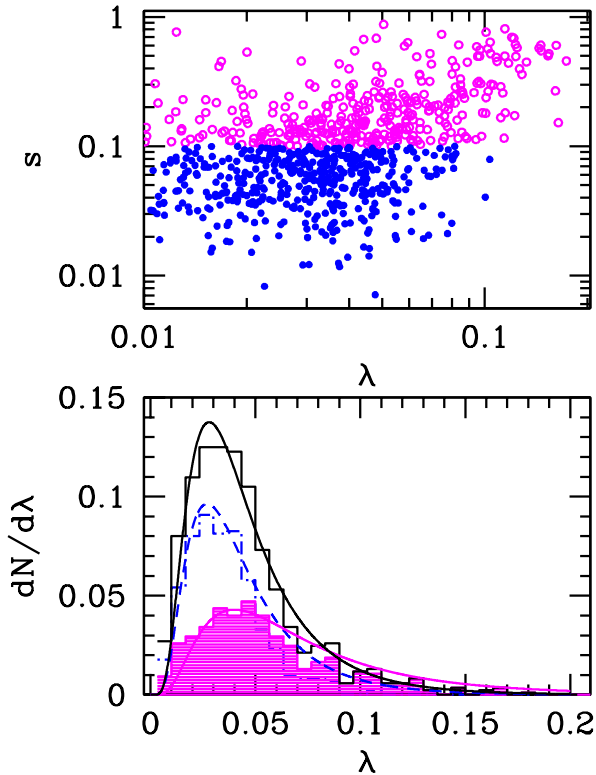}
\caption{{\it Top panel:} The off-center parameter, $s$, versus the
spin parameter, $\lambda$, measured within the virial radius at $z=0$
for all halos in our sample. Open circles correspond to ``unrelaxed''
systems ($s>0.1$); typically the remnants of recent accretion events
and mergers. Filled circles ($s<0.1$) denote ``relaxed'' systems
closer to virial equilibrium.  {\it Bottom panel:}  The distribution
of spin parameters of all halos (top histogram), ``relaxed'' halos
($s<0.1$, dot-dashed histogram), and ``unrelaxed'' systems ($s>0.1$,
shaded histogram). Curves show best lognormal fits to each histogram;
fit parameters are listed in the text.} \label{fig:svsl}
\end{figure}

Numerical simulations confirm this expectation: the spin distribution
of collapsed structures, as measured by the dimensionless spin
parameter, $\lambda=J |E|^{1/2}/GM^{5/2}$ ($J$ is the angular
momentum, $E$ is the binding energy, and $M$ is the total mass of a
halo), is wide and has a median value of $\lambda_{\rm med}\approx
0.035$ (see, e.g., Bullock et al. 2001, Gardner 2001, D'Onghia \&
Burkert 2004, Maccio' et al. 2006, Bett et al. 2007). The rather poor
efficiency of the spin-up process is also thought to explain the
extremely weak---but discernible---correlations between $\lambda$ and
other halo properties such as mass, formation time, and environment
(Cole \& Lacey 1996, Lemson \& Kauffmann 1999; Avila-Reese et
al. 2005; Allgood et al. 2006; Shaw et al. 2006; Hahn et al. 2007).

The tidal-torque scenario also implies that the specific angular
momentum acquired by the baryonic and dark matter components of a
galaxy should be similar, since the acquisition predates the collapse
of the system, during which non-linear effects may result in
substantial transfer of energy and angular momentum between the two
components (Navarro $\&$ Benz 1991; Navarro \& White 1994; Navarro \&
Steinmetz 1997, D'Onghia et al. 2006, Kaufmann et al. 2007). This
result underpins the assumptions of semianalytic work on galaxy
formation, where the angular momentum of a luminous galaxy is often
drawn randomly from the spin distribution of dark matter halos
obtained from cosmological N-body simulations (see, e.g., Cole et al
2000 and references therein).

Recently, however, a number of authors have highlighted the
possibility that mergers may play a substantial role in determining
the angular momentum content of a dark matter halo. D'Onghia \&
Burkert (2004), for example, argue that halos with a quiet merging
history might not acquire enough angular momentum to host late-type
spiral galaxies. Gardner (2001), Vitvitska et al. (2002), Peirani et
al. 2004, and Hetznecker \& Burkert (2006), among others, report a
significant correlation between mergers and spin and ascribe it to the
large orbital angular momentum associated with major
mergers. Vitvitska et al (2002), in particular, follow the evolution
of the spin parameter of the {\it most massive progenitor} of several
dark halos and find that the spin parameter varies with time,
increasing abruptly during major mergers, and decreasing gradually
during times of minor accretion.

This is an intriguing result, since late major mergers are normally
thought to be associated with the formation of elliptical galaxies
(see, e.g., the review by Burkert \& Naab 2003), which have long been
known to be---at fixed stellar mass---deficient in angular momentum
relative to spirals (Fall 1983). It is puzzling (and counterintuitive)
that galaxies where rotational support is low should tend to inhabit
dark halos where angular momentum is more plentiful.

We note, however, that the evolution of the spin of the most massive
progenitor of a halo might not be a faithful and direct measure of the
available angular momentum. Indeed, $\lambda$ is a {\it dimensionless}
rotation measure most useful for {\it isolated} systems (where $E$,
$J$, and $M$ are conserved), but of suspect applicability during a
major merger, when the identity of the system varies dramatically
with time.

On the other hand, angular momentum (total or specific) is a {\it
dimensional} quantity that typically increases as the mass and size of
a halo grow (see, e.g., Figure 8 of Navarro \& Steinmetz 1997). Late
major-merger remnants are, by definition, halos where a significant
amount of mass (and, in general, of angular momentum) has been
recently added to the halo. These late-assembling systems must have
``turned around'' later than an average halo of comparable mass. One
might therefore expect that tides have had a chance to operate for a
longer period of time, a fact that may explain their
higher-than-average spins.

Is this why merger remnants are reported to have higher-than-average
spins? Or do major merger remnants have instead spins comparable to
those of halos that have accreted a similar amount of mass on a
similar timescale through mainly smooth accretion? A proper answer to
these questions should evaluate the effect of mergers on halo spin
separately from the role of accretion, taking special care to define
the boundaries of the system so that spin measures are robust and
meaningful.

These are the issues that we address in this {\it Letter}.  In \S~2 we
present details of the numerical simulation and analysis procedure.
\S~3 discusses our main results, whilst \S~4 summarizes our
conclusions and implications.

\section{NUMERICAL METHODS}

\subsection{Simulations}

We analyze a cosmological N-body simulation of the $\Lambda$CDM
cosmogony, with cosmological parameters chosen to match the WMAP3
constraints (Spergel et al. 2006). These are characterized by the
present-day matter density parameter, $\Omega_0=0.238$; a cosmological
constant contribution, $\Omega_{\Lambda}$=0.762; and a Hubble
parameter $h=0.73$ ($H_0=100\, h$ km s$^{-1}$ Mpc$^{-1}$). The mass
perturbation spectrum has a spectral index, $n=0.951$, and is
normalized by the linear rms fluctuation on $8$ Mpc/$h$ radius
spheres, $\sigma_8=0.75$.

We simulate the evolution of $400^3$ particles of mass $m_{\rm{dm}}=
1.2 \times 10^{8}  \, h^{-1} \,M_{\odot}$ in a box 50 $h^{-1}$ Mpc
(comoving) on a side using the publicly available code GADGET2
(Springel 2005). Gravitational interactions between pairs of particles
are softened by a spline kernel with fixed comoving
Plummer--equivalent length of $3 \ h^{-1}$ kpc.

\subsection{Halo identification}
Non-linear structures at $z=0$ are identified using the classic
friends-of-friends (FOF) algorithm with a linking length equal to
$0.2$ times the mean interparticle separation.  For each FOF halo we
identify the most bound particle and adopt its position as the halo
center. Using this center, we compute the ``virial radius'' of each
halo, $r_{\rm vir}$, defined as the radius of a sphere of overdensity
$\Delta(z=0)=94$ (relative to the critical density for
closure)\footnote{The virial overdensity in a flat universe may be
computed using the fitting formula proposed by Bryan \& Norman (1998):
$\Delta (z)= 18 \pi^2 + 82\, f(z)-39\, f(z)^2$; with
$f(z)=\frac{\Omega_0(1+z)^3}{\Omega_0(1+z)^3+\Omega_{\Lambda}}-1$}. Quantities
measured within $r_{\rm vir}$ will be referred to as ``virial'', for
short. We select for our analysis all halos with masses in the range
$M_{\rm vir}=5 \times 10^{11}$ to $10^{13} \, h^{-1}$ M$_{\odot}$,
resulting in $848$ halos with $N_{\rm vir}$ betweeen $5,000$ and $80,000$
particles.

The assembly history of each halo may be studied using halo catalogs
analogous to the one just described, but constructed at various times
during the evolution of the system. For the purposes of our analysis,
we concentrate on the period $0<z<3$. Halos identified at $z>0$
are said to be {\it progenitors} of a $z=0$ system if at least $50\%$
of its particles are found within the virial radius of the
latter. Using this definition we can identify, at all times, the list
of progenitors of a given $z=0$ halo and track their properties
through time.

\begin{figure}
\includegraphics[width=85mm]{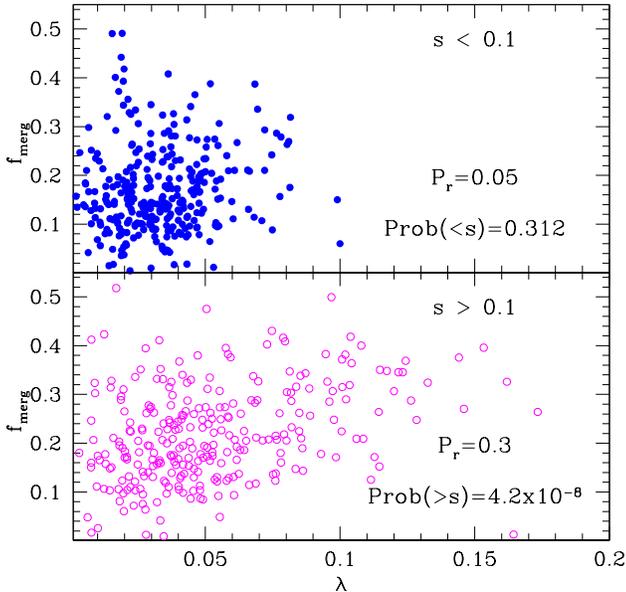}
\caption{The fraction of mass, $f_{\rm merg}$, accreted by a halo in
the single most important merger event since $z=3$, versus the spin
parameter, $\lambda$, measured within the virial radius at $z=0$ for
each halo in our sample. Top and bottom panels correspond to
``relaxed'' and ``unrelaxed'' halos, respectively.}
\label{fig:lvsfmrg}
\end{figure}

\subsection{Halo properties}

We compute, for all halos identified at $z=0$, the spin parameter
$\lambda$, using all particles within the virial radius. Note that the
choice of virial radius as a boundary defines implicitly the mass of
the halo and its angular momentum, and that the values of the spin
parameters assigned to a halo may be sensitive to this
choice. Especially sensitive will be halos that are ostensibly out of
equilibrium, since the virialization process will then lead to rapid
changes in the values of the virial radius, mass, and angular momentum
that define $\lambda$.

In order to assess these effects, we compute, for each halo, the
distance from the most bound particle to the center of mass of the
halo, normalized by the virial radius. This ``off-center'' parameter,
$s=|{\bf r}_{\rm{cm}}-{\bf r}_{\rm{mb}}|/r_{\rm{vir}}$, is a simple
but telling measure of the prevalence of unrelaxed substructure within
the halo and, consequently, of its equilibrium status. As discussed,
for example, by Hetznecker \& Burkert (2006), Maccio' et al. (2006),
and Bett et al (2007), halos with $s>0.1$ are remnants of relatively
recent accretion events which are still undergoing rapid changes in
their structural properties. Halos with $s<0.1$ are generally closer
to virial equilibrium and evolve weakly with time. Some exceptions do
occur, since ongoing mergers may by chance have (briefly) $s<0.1$,
but these exceptions are rare.

The importance of mergers during the assembly of the halo may be
estimated by the fraction of the mass contributed by the largest
accretion event in the assembly history of the halo, and is measured
in our analysis by the mass of the largest 2nd most massive progenitor
of a halo,
\begin{equation}
f_{\rm merg}=\frac{M_{\rm 2nd}(z<3)}{M_{\rm vir}(z=0)},
\end{equation}
identified since $z=3$, and normalized to the present-day virial mass
of the halo. This is a simple, albeit somewhat arbitrary, measure of
the importance of mergers, and we have verified explicitly that our
results are not overly sensitive to this choice. Varying the redshift
of identification of the 2nd most massive progenitor from $z=3$ to,
for example, the formation redshift of the halo (defined
as the time when the most massive progenitor first assembled half its
final mass) leads to no appreciable changes in our conclusions.

\section{RESULTS}

\subsection{Spin distribution and virial equilibrium}

The spin distribution of all halos selected at $z=0$ is shown by the
top histogram in the bottom panel of Figure~\ref{fig:svsl}, 
and may
be approximated by a lognormal distribution,
\begin{equation}
p(\lambda)\, \rm{d}\lambda=\frac{1}{\sigma_{\lambda}\sqrt{2\pi}}
\, \rm{exp}\Big[-\frac{\rm{ln}^2(\lambda/{\bar \lambda})}{2\sigma^2_{\lambda}}\Big] 
\frac{\rm{d}\lambda}{\lambda},
\end{equation}
with ${\bar \lambda} \sim 0.03$ and $\sigma_{\lambda}=0.58$,
consistent with previous work on the subject (Avila-Reese et al. 2005;
Allgood et al. 2006; Shaw et al. 2006; Maccio' et al. 2006; Bett et
al. 2007, Hernandez et al. 2007). This fit is shown by the thick solid
line in the same panel.

The other two histograms in the bottom panel of Figure~\ref{fig:svsl}
correspond to splitting the halo sample according to the off-center
parameter $s$. ``Unrelaxed'' halos (i.e., those with $s>0.1$, shaded
histogram) make up $46\%$ of the sample, and tend to have
higher-than-average spins. ``Relaxed'' halos (i.e., those with
$s<0.1$; dot-dashed histogram) have a narrower $\lambda$ distribution
more sharply peaked around slower rotators. This difference is
reflected in the parameters of the best lognormal fits to each
histogram, which give ${\bar \lambda}$=0.028 and
$\sigma_{\lambda}=0.58$ for relaxed halos, and ${\bar \lambda}$=0.04
and $\sigma_{\lambda}=0.65$ for unrelaxed systems.

The difference in the spin of relaxed and unrelaxed halos is
responsible for the clear trend between $s$ and $\lambda$ seen in the
top panel of Figure~\ref{fig:svsl}. The majority of slowly-rotating
halos are in equilibrium: $71\%$ of halos with $\lambda<0.02$ have
$s<0.1$.  This fraction drops to $59\%$ for halos with
$0.03<\lambda<0.06$, and there are basically no equilibrium halos with
$\lambda>0.1$. Since the remnants of recent major mergers undoubtedly
figure prominently in the unrelaxed halo sample, the trend apparent in
the top panel of Figure~\ref{fig:svsl} is reminiscent of the findings
of Gardner (2001) and Vitvitska et al. (2002) referred to in
\S~\ref{sec:intro}.

\subsection{Mergers and spin}

The relation between mergers and spin is explored in
Figure~\ref{fig:lvsfmrg}, where the top panel shows, for relaxed
($s<0.1$) halos, the dependence of the spin parameter on $f_{\rm
merg}$, the fraction of mass accreted by a halo in its single most
important merger event since $z=3$. This panel shows that there is no
obvious correlation between $\lambda$ and merging activity when
considering systems near virial equilibrium.

The visual impression is confirmed by statistical measures of
association applied to this sample. For example, the Pearson's linear
coefficient correlation of $s<0.1$ halos is $P_r=0.05$, with less than
$31\%$ probability that the null hypothesis of zero correlation may be
disproved. Varying our adoption of $s=0.1$ as the definition of
relaxed and unrelaxed halos has little effect on this conclusion, as
may be seen from Table 1: choosing $s<0.075$ or $s<0.125$ to define
``relaxed'' halos also results in sizable probabilities, from $65\%$
to $7\%$, respectively.

\begin{table*}
\caption{\hspace{5.5cm} Parameters of the linear correlation coefficient}
\vspace{0.3cm}

\begin{tabular}{c|c|c|c|c|c|c|c} \hline
s & N($<$s) & $P_r$($<$s) & Prob($<$s) & N($>$s) & $P_r$($>$s) &
Prob($>$s) & \% halos($<$s) \\ \hline \hline

0.125  &  567  &   0.08    &  0.071  &  281  &  0.25  &  $7.1\rm{x}10^{-5}$  & 66.8\% \\
0.1    &  457  &   0.05    &  0.312  &  391  &  0.30  &  $4.2\rm{x}10^{-8}$  & 54\% \\
0.075  &  354  &   0.025   &  0.657  &  494  &  0.32  &  $7.2\rm{x}10^{-12}$ & 41.8\% \\
\hline

\hline
\end{tabular}\\
\end{table*}

Note that the $s<0.1$ relaxed halo sample includes many remnants of
major merger events; $13\%$ of them have accreted more than $25\%$ of
their mass in a single merger, and half of them have undergone at
least one merger during which more than $\sim 15\%$ of their final
mass was accreted. Yet, there is no obvious evidence that more massive
mergers lead to higher than average spin parameters. {\it We conclude
that merging history and spin are uncorrelated in equilibrium halos.}

\begin{figure}
\centering\includegraphics[width=1.0\linewidth,clip]{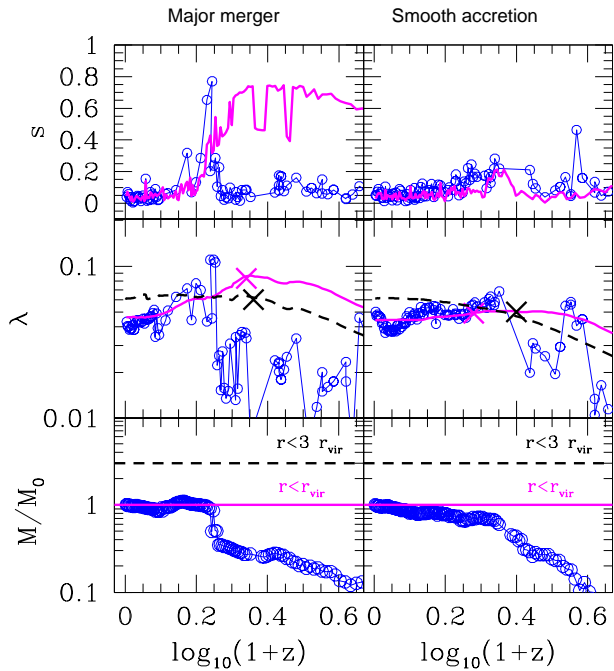}
\caption{Evolution of the off-center parameter, $s$, the spin
parameter, $\lambda$, and the mass, $M$, for two halos chosen to have
a similar value of $\lambda$ and to be in equilibrium at $z=0$. Masses
are scaled to the virial mass of each halo at $z=0$. The halos differ
in their merging history. One of the halos (left panels) undergoes a
major merger at $z=0.7$ ($f_{\rm merg}=0.36$), whereas the other
(right panels) has a smoother accretion history ($f_{\rm
merg}=0.07$). Open circles correspond to the most massive progenitor
of the halo, whereas solid and dashed lines correspond, respectively,
to {\it all} particles identified within one and three times the virial radius at
$z=0$.}\label{fig:sev}
\end{figure}

The situation is different for ``unrelaxed'' halos, as shown in the
bottom panel of Figure~\ref{fig:lvsfmrg}. Here a correlation between
$f_{\rm merg}$ and $\lambda$ is clearly present, a result that is
statistically significant according to the linear correlation analysis
summarized in Table 1.  In particular, there is a tail of high-spin
($\lambda>0.1$) halos that is almost exclusively populated by systems
that have added at least $\sim 30\%$ of their mass in relatively
recent merger events.

Because these systems are out of equilibrium, spin estimates are
likely to fluctuate as they relax, mostly as a result of the evolving
boundaries of the system imposed by the virial definitions discussed
in \S 2.2. These fluctuations must lead to a net overall reduction in
the spin measured within the virial radius in order to explain the
difference in the spin distribution of relaxed and unrelaxed halos.

\subsection{Spin evolution}

We explore this in Figure~\ref{fig:sev}, where we follow the evolution
of two halos selected to have similar spin ($\lambda=0.045$) and to be
in equilibrium (i.e., $s<0.1$) at $z=0$, but of very different merging
histories. One of them (shown in the left panels of
Figure~\ref{fig:sev}) has seen a large fraction of its mass accreted
in a major accretion event ($f_{\rm merg}=0.36$), whereas the other
(right-hand panels) has had a much smoother accretion history ($f_{\rm
merg}=0.07$). 

Open circles in this figure correspond to the most massive progenitor
identified at each redshift. In the case of the major merger remnant,
its mass more than doubles between $z\sim 0.8$ and $z\sim
0.7$. Subsequent evolution adds little extra mass to the system, which
gradually relaxes into equilibrium: the center offset parameter, $s$,
drops from a maximum of almost unity at the height of the merger
($z\sim 0.75$) to $s<0.1$ by $z\sim 0.3$, and it remains in that range
until the present.

For this halo, the evolution of the spin parameter of the most massive
progenitor evokes the results of Vitvitska et al (2002): $\lambda$
rises during the merger to $\lambda\sim 0.1$ and then drops gradually
to its final value of $\lambda=0.045$ at $z=0$. Since energy and
angular momentum are conserved in collisionless mergers, the net
reduction in spin must occur as a result of the internal
redistribution of mass and angular momentum ensuing the merger. This
process tends to populate the central regions of the remnant with
low-angular momentum material and to push high angular momentum
particles into weakly bound orbits. As a result, the spin measured
{\it within} the (evolving) virial radius of the halo steadily drops
although the {\it total} energy and angular momentum of the system is
conserved.

This may be seen by following the spin parameter measured over a
larger volume that includes the halo. The solid and dashed lines
without symbols in Figure~\ref{fig:sev} correspond, respectively, to
{\it all} particles that end up, at $z=0$, within one and three virial
radii from the center of the halo. The comparison between these two
curves shows that, when measured over an expanded region not subject to
arbitrary ``virial'' boundaries, the time evolution of the spin
parameter follows closely the behaviour expected from tidal torque
theory: net spin is acquired early and evolves little after
turnaround{\footnote{Turnaround is defined as the time when the moment of
inertia of each subset of particles peaks and is marked by a cross on
each curve.}}.

The dashed line in the mid-panel of Figure~\ref{fig:sev}, in
particular, shows that the spin of the larger region where the halo is
embedded is acquired well before the merger takes place, and remains
constant after turnaround despite the major merger event at
$z=0.73$. The merger-induced spin-up and subsequent spin-down of the
most massive progenitor during a merger is thus clearly a result of
restricting the computation of mass, energy, and angular momentum to a
subset of the system, namely that contained within the evolving virial
radius of the halo---i.e., the most massive progenitor.  Although we
have chosen a particular halo for illustration in the left panels of
Figure~\ref{fig:sev}, we have explicitly checked that similar results
are consistently found when applying this analysis to all major merger
remnants.

Similar results are found for systems that have been assembled through
smoother accretion merging histories (right-hand panels in
Figure~\ref{fig:sev}) Here again, and despite the large differences in
merging activity, the spin parameter of the region where the halo is
embedded is fixed at turnaround and remains approximately constant
thereafter. Interestingly, the after-merger spin-down of the most
massive progenitor is not obvious here, implying that the
redistribution of angular momentum from the center outwards
responsible for that drop is most efficient during major mergers.

\section{Summary}

We have used numerical simulations to examine the relation between the
merging history and spin of cold dark matter halos.  Our results show
that, as expected from standard tidal torque theory, the angular
momentum of the material destined to collapse into a halo is acquired
during the early expansion phase and evolves little after turnaround,
regardless of whether the subsequent assembly of the halo involves
mainly major mergers or smooth accretion. Major mergers, {\it per se},
do not result in higher-than-average spins: we fail to find a
significant correlation between merging history and halo spin in
equilibrium halos.

Unrelaxed halos, on the other hand, do have a spin distribution that
peaks at higher values than equilibrium systems.  Since these halos
must eventually relax, this implies that virialization leads in
general to a net reduction in the spin of a system. We show that this
results from the redistribution of matter and angular momentum that
accompanies virialization, a process that may push high angular
momentum material outside the virial radius of a halo and that is
especially efficient during major mergers.

These results call into question the common practice of assuming that
there is a tight correspondence between the angular momentum of a halo
(measured within its virial radius and without regard for its
equilibrium status) and that of its baryonic component. Baryons and
dark matter will be affected differently by the redistribution of
angular momentum that occurs during virialization, depending largely
on the spatial segregation of one component relative to the other.
The more segregated the baryons are before the assembly of the system
is completed, the more angular momentum they will tend to transfer to
the dark matter as they spiral to the center to form the final system.

The efficiency of this process is likely to vary strongly with merger
history; to affect most prominently the remnants of recent major
mergers; and to result, overall, in a rather complex relation between
the angular momentum of a halo and that available to its luminous,
baryonic component. Only simulations that follow, in a realistic
manner, the cooling and condensation of the baryons within the
multiple stages of the hierarchical assembly of a galaxy system are
likely to capture the true interdependence between spin and assembly
history of a galaxy.

\section*{Acknowledgements} 
E.D. is supported by a EU Marie Curie Intra-European Fellowship under
contract MEIF-041569. The numerical simulations were performed a the
SGI Altix system of the University Sternwarte in Munich. E.D. thanks
Tobias Kaufmann and Justin Read for stimulating discussions. JFN
acknowledges support from the Alexander von Humboldt Stiftung and from
the Leverhulme Trust, as well as the hospitality of the Institute for
Computational Comology at the University of Durham.

\label{lastpage}
\end{document}